# Recommending Low-Cost Compact Space Environment and Space Weather Effects Sensor Suites for NASA Missions

Yihua Zheng[1], Michael A. Xapsos[2], Insoo Jun[3], T. P. O'Brien[4], Linda Parker[5], Wousik Kim[3], Justin Likar[6], Joseph Minow[7], Thomas Y. Chen[8], Douglas Rowland[9]

[1] NASA Goddard Space Flight Center, Space Weather Laboratory, Greenbelt, MD, USA.
[2] NASA Goddard Space Flight Center, Radiation Effects and Analysis Group, Greenbelt, MD, USA.
[3] Mission Environments Group, Jet Propulsion Laboratory, California Institute of Technology, Pasadena, CA, USA.
[4] Aerospace Corporation, Chantilly, Virginia, USA.
[5] Space Weather Solutions, Alabama, USA.
[6] Johns Hopkins Applied Physics Laboratory, USA.
[7] NASA Technical Fellow for Space Environments, Langley Research Center, Hampton, VA, USA.
[8] Columbia University, USA.
[9] NASA Goddard Space Flight Center, The Ionosphere, Thermosphere, Mesosphere Physics Laboratory, Greenbelt, MD, USA.

**Synopsis**:
As miniaturized spacecraft (e.g., cubesats and smallsats) and instrumentation become an increasingly indispensable part of space exploration and scientific investigations, it is important to understand their potential susceptibility to space weather impacts resulting from the sometimes volatile space environment. There are a multitude of complexities involved in how space environment interacts with different space hardware/electronics. Measurements of such impacts, however, have been lacking. Therefore, we recommend developing and/or procuring low-cost, low-power consumption, and compact sensor suites (mainly for space weather and impact purposes) and flying them on all future NASA (and, more generally, U.S) missions in order to measure and quantify space weather impacts, in addition to the main instrumentation.

**Synopsis**:

As miniaturized spacecraft (e.g., cubesats and smallsats) and instrumentation become an increasingly indispensable part of space exploration and scientific investigations, it is important to understand their potential susceptibility to space weather impacts resulting from the sometimes volatile space environment. There are multitude of complexities involved in how space environment interacts with different space hardware/electronics. Measurements of such impacts, however, have been lacking. Therefore, we recommend developing and/or procuring low-cost, low-power consumption, and compact sensor suites (mainly for space weather and impact purposes) and flying them on all future NASA (and U.S in general) missions in order to measure and quantify space weather impacts, in addition to the main instrumentation.

1. **Motivation**

The current trend towards miniaturization of spacecraft (e.g., smallsats and cubesats) and instrumentation is expected to continue, with such hardware likely to play an increasingly significant role in future space exploration activities as well as in scientific investigations. Such missions are vulnerable to the effects of space weather. Indeed, environmental effects have become more complex, problematic, and at times destructive. A better understanding of how miniaturized space hardware/microelectronics fares in a dynamically varying space environment is becoming critical. Yet there is a paucity of such measurements, in part due to their secondary role in many space missions, with financial constraints often coming into play.

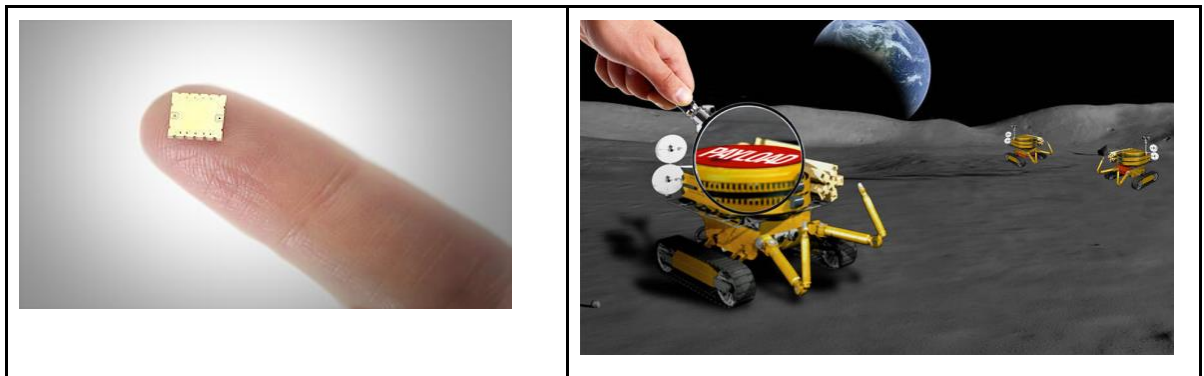

*Fig1: Miniaturization of space hardware/space assets*

The SEESAW (Space Environment Engineering and Science Applications) Roadmap Report (**https://cpaess.ucar.edu/sites/default/files/2022-06/atr-2019-00431-seesaw-sorkshop-roadmaps.pdf**) draws attention to the need for accurate measurements. It also emphasizes that 'not only environmental measurements are needed, but also sensors and missions dedicated to measuring anomalies are needed' in order to correctly connect space environment to observed satellite anomalies.

2. **Recommendations:**

- We recommend developing/procuring low-cost, low-power, and compact sensors and flying them (e.g., via hosted payloads, or dedicated cubesat/smallsat missions) whenever/wherever possible to better characterize the environment and to better understand how the space environment interacts with space electronics/hardware.
- We recommend an open data policy once the measurements are collected and verified. We advocate best and full utilization of data by involving relevant experts (instrumentalist, space environment experts, engineering impacts experts, end users, and so on) and through international coordination and collaborations.

**Potential Benefits:**

- A better understanding of space environment and its interactions with space assets can also lead to possible cost reductions by incorporating an appropriate level of radiation hardening and charging mitigation, and by avoiding over-protection in space hardware design (primary).
- The potential to assist with anomaly resolution and risk mitigation; a more robust understanding of how parts and materials respond/degrade over their mission time and in different regions of space (primary)
- Advancement of scientific knowledge of space environment (secondary)

## 3. Space Environment/Effects Measurements: Status

Japan

In recognition of the importance of space environment measurements (high energy particles, galactic cosmic rays (GCRs), atomic oxygen, plasma, and direct effects) on parts and materials, JAXA has mounted measuring instruments onboard each spacecraft since ETS-V (Engineering Test Satellite V) was launched in 1987.

JAXA has built the 'Space Environment and Effects System' (https://sees.tksc.jaxa.jp/fw_e/dfw/SEES/English/Top/top_e.shtml).

Types of Instruments can be reviewed via the link below: https://sees.tksc.jaxa.jp/fw_e/dfw/SEES/English/Instruments/instruments_e.shtml#DOM

Engineering effects focused ones (examples) include

- DOM: DOse Monitor / SDOM: Standard DOse Monitor
- DOS: DOSimeter
- AOM: Atomic Oxygen Monitor
- POM: POtential Monitor
- DIM: DIscharge Monitor
- SUM: Single event Upset Monitor / RSM: RAM Soft-error Monitor
- ICM: Integrated Circuit Monitor
- SCM: Solar Cell Monitor
- COM: COntamination Monitor / TDM: Thermal control material Degradation Monitor

**Europe**

In Europe, space environment and effects sensors flown in space include SURF (carried on the STRV1d GTO mission), which provides measurement of electron current (internal charging); the Merlin space weather monitor (Ryden et al., 2004), which measures the internal charging, total ionizing dose, proton fluxes, and ion LET (Linear Energy Transfer) on Galileo Giove-A; and the CEDEX (Cosmic Ray Energy Deposition Experiment) on Giove-A. Note that CREDANCE in the next Section builds upon SURF, Merlin and CEDEX, with a heavy ion sensor added. Others include SREM (Standard Radiation Environment Monitor) on Giove-B (measuring >0.5 MeV electrons, >10 MeV protons, and >150 MeV ions). Recently, the European Space Agency (ESA) has initiated the D3S (Distributed Space Weather Sensor System) effort, with the goal of providing space weather measurements as an input to future space weather operational services. NGRM (Next Generation Radiation Monitor) sensors (with SREM as the predecessor), as part of the D3S, have been placed on the EDRS-C (European Data Relay System, Satellite-C) at GEO) since 2019 and the Sentinel-6 Michael Freilich (S-6-MF) satellite (at LEO) since 2020 (e.g., Sandberg et al., 2022). Such space weather sensors will be continuously flown on future ESA missions.

**U.S.**

Direct space environment effects measurements have been limited. The few dedicated missions include SCATHA (Spacecraft Charging at High Altitudes), which was launched on 1/30/1979. Its measurements have formed the major knowledge base regarding spacecraft charging anomalies and their space environment connections that we still rely on today, Other examples include CRRES (Combined Release and Radiation Effects Satellite; mission from July 25, 1990 to October 12, 1991), whose measurements include the radiation environment of the inner and outer radiation belts, and radiation effects on microelectronics devices. Space Environment Testbeds (SET) onboard DSX (Demonstration and Science eXperiments) (G. Spanjers et al., 2006; launched on 6/25/19) included several instruments measuring direct space weather impacts; these have proven to be useful in providing critical information needed on how space hardware and electronics interact with the space environment and also for testing out different detector technology/functions (e.g., Poole, McNulty, and Poole, 2022). As an auxiliary instrument, the Engineering Radiation Monitor (ERM) measures dose, dose rate, and charging currents on the Van Allen Probes mission. Such data have provided a dynamic response of host satellites/instruments to the radiation environment and instrument performance over time. Microdosimeters, such as the two inside RPS (Relativistic Proton Spectrometer) on board Van Allen Probes (the second generation of those first flown on the Lunar Reconnaissance Orbiter; Mazur et al., 2011), are useful to have with their low mass, low power, and a small footprint.

We are in need of **more direct space environment effects measurements. at different orbits/regions (including at Moon, Mars, and beyond).** These compact and low-cost sensors make local environment measurements possible that depend on the local geometry (shielding and sunlight intensity) and temperature. These local environment measurements can be directly used for anomaly resolution and for situational awareness.

## 4. Potential Compact Sensors for NASA Missions

Potential sensors can be similar to those on board DSX and the others mentioned above. However, they can be even more compact, lower-cost, and consuming less power. Below are examples of sensors onboard DSX.

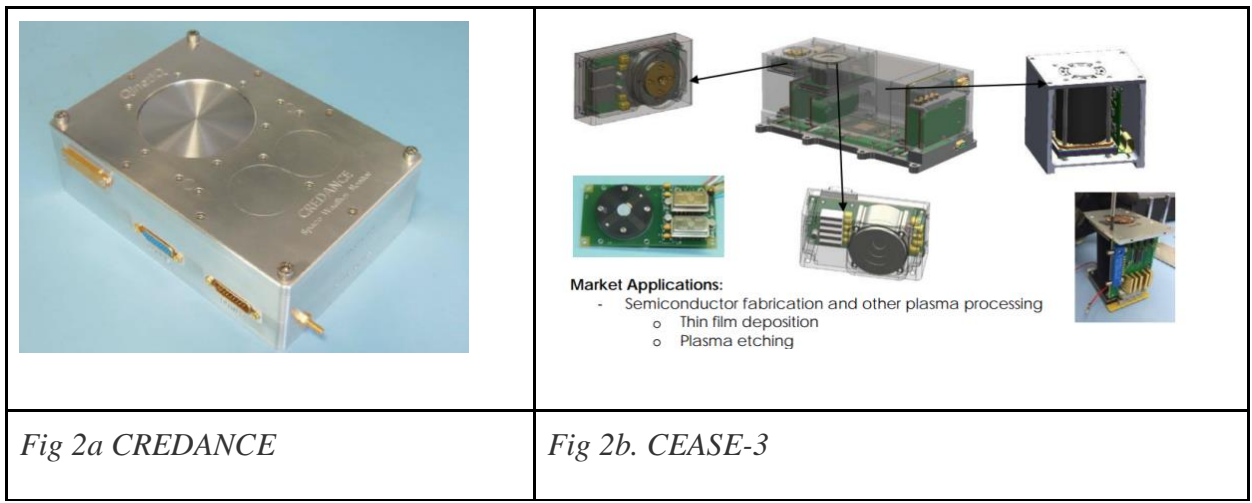

| *Fig 2a CREDANCE* | *Fig 2b. CEASE-3* |

*CREDANCE (Cosmic Radiation Environment Dosimetry and Charging Experiment)*

This is a third-generation sensor suite consisting of a proton and heavy ion telescope. It provides internal charging current measurements at three shielding depths, and total ionizing dose measurements at two shielding depths. CREDANCE is fabricated at Surrey Space Center, University of Surrey, UK.

Measurement Specs are: > 40 MeV proton flux; 0.1 to 28.5 MeV-cm2/mg heavy ion LET; internal charging current at 0.5, 1.0 and 1.5 mm Al (aluminum) shielding; ionizing dose at 2.5 and 5.5 mm Al shielding.

Size and Power Specs: 18.5 x 12.5 x 6 cm, 1.7 kg, 2.4 W.

*CEASE3 (Compact Environmental Anomaly Sensor).*

This is a third-generation sensor suite consisting of low, medium and high energy electron and proton telescopes and an electrostatic analyzer.

- Measurement Specs: 0.1 keV to 5 MeV electrons; 2 to 100 MeV protons.
- Size and Power Specs: 26.9 x 11.1 x 10.4 cm; 3.9 kg; 12.5 W.

Other types of sensors onboard DSX include:

*DIME (Dosimetry Intercomparison and Miniaturization Experiment)*

Dosimetry Intercomparison and Miniaturization (DIME) measures space radiation environments that are detrimental to space system reliability using novel dosimetry techniques. DIME occupies two 3U card slots on the SET-1 carrier.

*ELDRS (Enhanced Low Dose Rate Sensitivity) Instrument*

ELDRS is the enhancement of degradation in transistors or circuits when exposed to radiation at low dose rates as compared to high dose rates (e.g., Benedetto et al., 2021). ELDRS has introduced new challenges for radiation hardness assurance. The ELDRS instrument on DSX performs real-time, in-orbit measurements of radiation-induced degradation from trapped particles by monitoring changes in the collector and base currents of active BJTs (Bipolar Junction Transistors).

*COTS-2*

COTS-2 measures particle-induced single event effects on an SEE (single event effects) mitigation platform in normal backgrounds and during solar storm events. COTS-2 will measure a SEE on COTS (Commercial Off the Shelf) FPGAs (Field Programmable Gate Arrays), classify the event by event type, and determine if mitigation of the effect occurred without watchdog intervention.

*New-Generation Sensors*

A new generation of sensors that take advantage of technology advances and innovation should be developed and procured. At Aerospace Corporation, modular compact sensors aimed at build-to-order and at COTS prices have been developed or are in the pipeline. These sensors can be flown individually or combined into an all-hazards suite sensor. However, for some of the sensors, manufacturing and integration remain costly. Hopefully, such challenges will be resolved with further sensor development and collaboration.

*Space radiation/environment monitors at Moon and Mars*

Two examples currently in orbit are the RAD (Radiation Assessment Detector) onboard MSL (Mars Science Laboratory) and the Cosmic Ray Telescope for the Effects of Radiation (CRaTER) onboard LRO. Measurements from both detectors have been crucial for radiation environment monitoring at their respective locations.

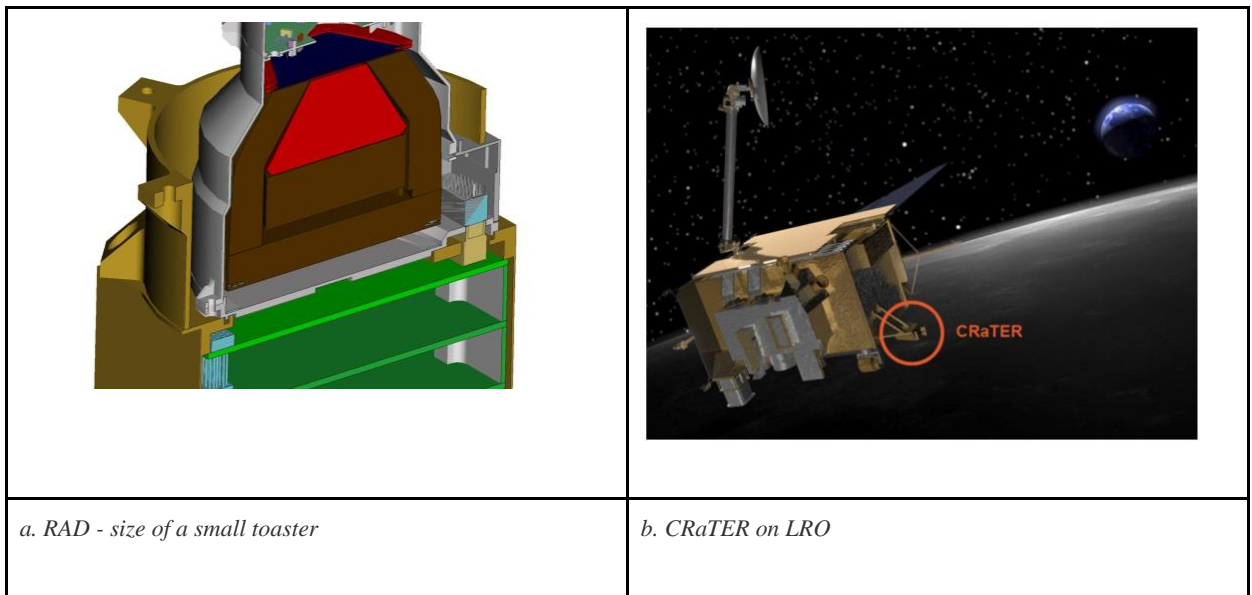

| a. RAD - size of a small toaster | b. CRaTER on LRO |

*Figure 3. Radiation Monitors flown currently*

The HERMES (Heliophysics Environmental and Radiation Measurement Experiment Suite) mission and ERSA (European Radiation Sensors Array) to be mounted on Lunar Gateway will provide critical measurements for both science and space weather monitoring purposes, crucial for future human missions to the Moon and Mars.

5. **SUMMARY**

In summary, space environment and effects sensors should be mounted on all future NASA missions; this will help better define the complex effects and mechanisms of the space environment; reduce design margins; foster increased use of space environment "tolerant" technologies; increase the fraction of payload resources; and reduce launch vehicle requirements. The measurements from such sensors will lead to the improvement of design and operations guidelines. In addition, such data will help with science understanding and advancement. Any development, procurement, and deployment of cost-effective, reliable, innovative, and high-quality sensors should be encouraged, and not just limited to the examples shown above. Once observations are made, another critical element is to have efficient data management, processing, distribution, and utilization plans to increase the value of such assets.

**References:**

Ryden, Keith & Dyer, C. & Morris, P.A. & Haine, R.A. & Jason, Susan. (2004). The merlin space weather monitor and its planned flight on the Galileo System Testbed Satellite (GSTB-V2/A). 12. 8085-8092.

G. Spanjers et al., "The AFRL demonstration and science experiments (DSX) for DoD space capability in the MEO," 2006 IEEE Aerospace Conference, 2006, pp. 10 pp.-, doi: 10.1109/AERO.2006.1655750.

I. Sandberg *et al*., "First Results and Analysis From ESA Next Generation Radiation Monitor Unit Onboard EDRS-C," in *IEEE Transactions on Nuclear Science*, vol. 69, no. 7, pp. 1549-1556, July 2022, doi: 10.1109/TNS.2022.3160108.

A. R. Benedetto, H. J. Barnaby, C. Cook, M. J. Campola and A. Tender, "BJTs in Space: ELDRS Experiment on NASA Space Environment Testbed," *2021 IEEE Radiation Effects Data Workshop (REDW)*, 2021, pp. 1-5, doi: 10.1109/NSREC45046.2021.9679350.

SEESAW Roadmap Report: **https://cpaess.ucar.edu/sites/default/files/2022-06/atr-2019-00431-seesaw-sorkshop-roadmaps.pdf**